\documentclass[
reprint,
superscriptaddress,
amsmath,
amssymb,
aps,
prl
]{revtex4-2}

\usepackage[T1]{fontenc}
\usepackage[utf8]{inputenc}
\usepackage{graphicx}
\usepackage{xcolor}
\usepackage{bm}
\usepackage{dcolumn}
\usepackage{hyperref}
\usepackage{cleveref}
\crefname{figure}{Fig.}{Figs.}
\Crefname{figure}{Fig.}{Figs.}
\crefname{equation}{Eq.}{Eqs.}
\Crefname{equation}{Eq.}{Eqs.}
\hypersetup{
	colorlinks=true,
	linkcolor=blue,
	citecolor=blue,
	urlcolor=blue
}

\begin{document}
\hyphenpenalty=10000
\exhyphenpenalty=10000

\title{Local Electroneutrality Violation as a Universal Constraint in Confined Electrolytes}

\author{Marcelo Lozada-Cassou}
\email{marcelolcmx@ier.unam.mx}
\affiliation{Instituto de Energ\'ias Renovables, Universidad Nacional Aut\'onoma de M\'exico, Temixco, Morelos 62580, Mexico}

\date{\today}

\begin{abstract}
	We show that finite-size violations of local electroneutrality in confined electrolytes are governed by the topology of the confining domain, yielding a universal hierarchy of deviations across spherical, cylindrical, and planar geometries. Within Poisson–Boltzmann theory, we introduce an electroneutrality deviation ratio that quantifies how global electrostatic constraints associated with compactness and boundary multiplicity modify charge balance inside confined domains. Although electroneutrality is asymptotically restored in all geometries, finite-size deviations are strongest in compact spherical cavities, weaker in cylindrical confinement, and weakest in planar slits. These results identify topology as the structural origin of confinement-induced charge redistribution and establish the violation of local electroneutrality as a global constraint underlying phenomena such as overcharging and charge reversal, demonstrating that confinement—not local geometric detail—controls the emergence of these effects.
\end{abstract}
\maketitle
\textit{Introduction.—}
Confinement profoundly modifies the electrostatic and thermodynamic behavior of electrolyte systems. When charged fluids are restricted to finite domains such as nanopores, hollow nanoparticles, or slit geometries, the presence of boundaries induces charge redistribution and deviations from local electroneutrality~\cite{Lozada_1984,Luo-electroneutrality-nature-2015,Levin_electroneutrality-2016,Levy-electroneutrality-2020,Levy-electroneutrality-PRE-2021,Green-electroneutrality-JCP-2021}. These effects are typically described in terms of geometric properties such as curvature or confinement length scales. However, despite extensive studies of confined electrolytes, a unifying principle that explains how global structural features of the domain control these phenomena remains lacking. Previous studies have reported confinement-induced charge redistribution~\cite{Adrian-JML-2023,Lozada-Cassou_JML-2025} and symmetry breaking~\cite{Lozada-Cassou_Symmetry_breaking_2025} in specific geometries.

In this Letter, we show that the finite-size violation of local electroneutrality in confined electrolytes is controlled by the topology of the confining domain. By analyzing spherical, cylindrical, and planar geometries within Poisson--Boltzmann theory~\cite{Verwey_TheoryStabilityLyophobicColloids_1948,Andelman-2006}, we identify a universal hierarchy of electroneutrality deviations governed by the compactness and connectivity of the domain. These results demonstrate that confinement-induced charge redistribution is governed by global electrostatic constraints imposed by topology, rather than by local geometric properties.

The central idea is that confinement-induced electrostatic and thermodynamic effects are structurally determined by the topology of the confining domain rather than by its detailed geometric realization. Hollow charged nanoparticles can be naturally described as product manifolds of the form
\begin{equation}
	\Omega \simeq \mathcal{M} \times [0,\delta],
\end{equation}
where $\mathcal{M}$ denotes the mid-surface of the shell (e.g., $\mathcal{M}=S^{2}$ for a spherical shell, $\mathcal{M}=S^{1}\times[0,L]$ for a finite cylindrical shell, or $\mathcal{M}=S^{1}\times\mathbb{R}$ for an infinite cylinder), and $[0,\delta]$ represents the transverse thickness. Within this representation, curvature acts as a geometric refinement, whereas the existence and structure of global constraints are determined by the topology of $\mathcal{M}$ and its boundary components.

\textit{Discussion.—}At the mean-field level, the electrostatic potential of a cavity shell immersed in an electrolyte, satisfies the Poisson--Boltzmann equation~\cite{Verwey_TheoryStabilityLyophobicColloids_1948}
\begin{equation}
	-\varepsilon \nabla^{2} \psi(\mathbf{r}) 
	= \rho_{f}(\mathbf{r}) 
	+ \sum_{i} z_{i} e\, n_{i}^{0}
	\exp\!\left[-\beta z_{i} e \psi(\mathbf{r})\right],
	\label{eq:PB}
 \end{equation}

where $\varepsilon$ is the dielectric permittivity, $\rho_f$ the fixed charge density, $z_i$ the ionic valence, $n_i^0$ the bulk concentration, $e$ the elementary charge, $ \psi(\mathbf{r})$ is the mean electrostatic potential, $T$ is the temperature, and $\beta=(k_BT)^{-1}$. Solving \Cref{eq:PB}, the local charge density inside and outside the cavity shell, $\rho_{el}(\mathbf{r})= \sum_{i} z_{i} e\, n_{i}^{0}
\exp\!\left[-\beta z_{i} e \psi(\mathbf{r})\right]$, is obtained, where $\mathbf{r}$ is measured from its geometrical center. 

To quantify finite-size deviations from local electroneutrality under confinement, we define the electroneutrality deviation ratio

\begin{equation}
	\eta(R)=\frac{Q_{\mathrm{in}}(R)+Q_{\mathrm{in}}^{0}}{Q_{\mathrm{in}}^{0}},
\end{equation}

where $Q_{\mathrm{in}}(R)=\int_{\omega_{\mathrm{in}}} \rho_{el}(\mathbf{r})\, dV$ is the total mobile charge within the cavity of characteristic size $R$, and $Q_{\mathrm{in}}^{0}$ is the total fixed charge on the inside confining boundary. A nonzero value of $\eta(R)$ signals a violation of local electroneutrality, hereafter referred to as the violation of the local electroneutrality condition (VLEC).

The electrostatic potential is determined self-consistently by the global charge distribution, which couples the interior and exterior regions of the confined domain through long-range Coulomb interactions~\cite{Lozada-Cassou-PRE1997}. As a result, the charge distribution inside the cavity cannot be determined independently of the exterior region, even at the mean-field level.

This definition enables a direct comparison of electroneutrality deviations across geometries with distinct topological properties.

\begin{figure}[htbp]
	\includegraphics[angle=360,width=1.1\linewidth]{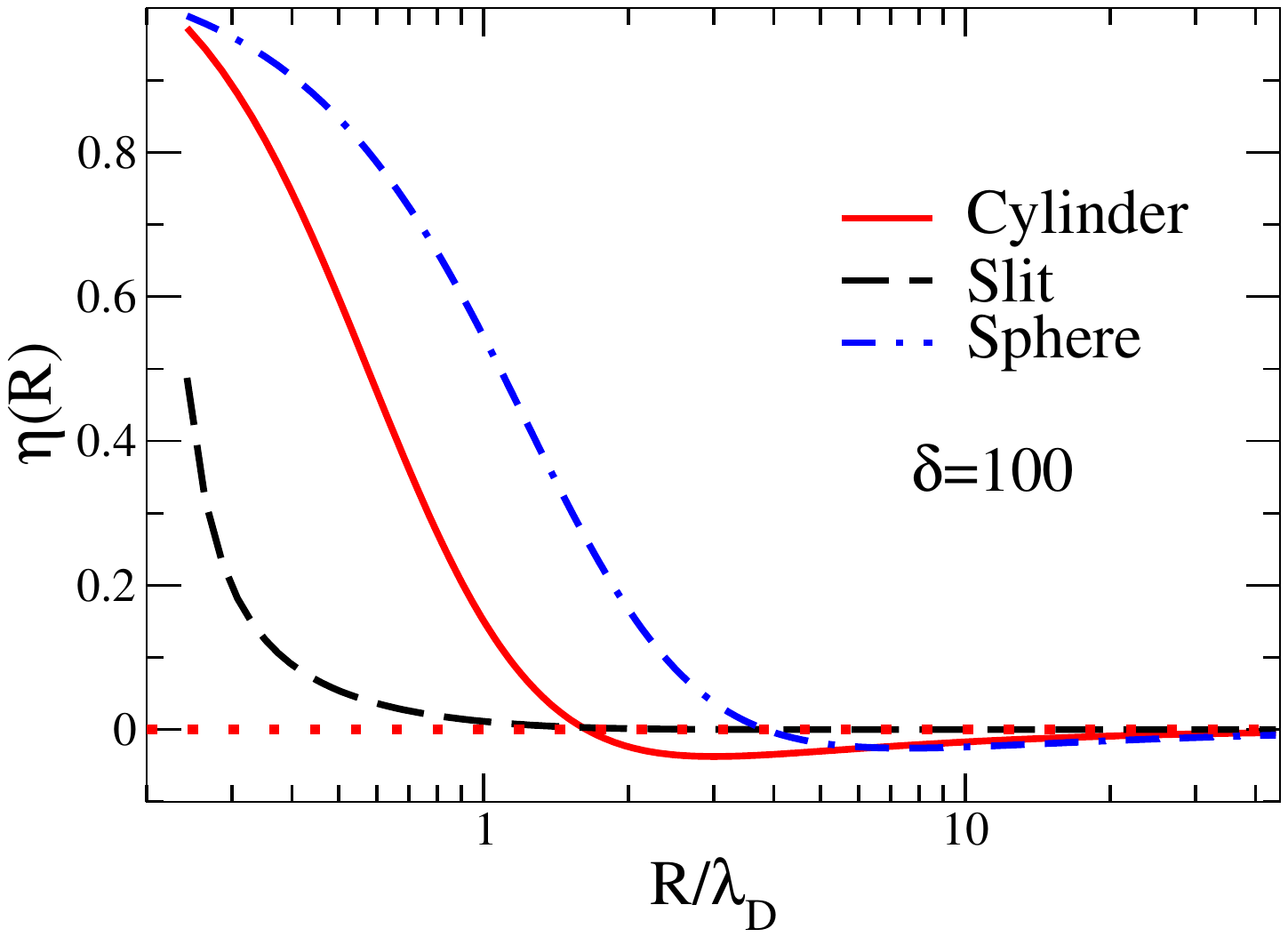}
	\caption{
		Electroneutrality deviation ratio $\eta(R)=(\sigma_{Hi}(R)+\sigma_{0})/\sigma_{0}$ for planar slits ($\Omega_{\mathrm{slit}}\simeq \mathbb{R}^{2}\times[0,\delta]$), infinite cylindrical shells ($\Omega\simeq S^{1}\times\mathbb{R}\times[0,\delta]$), and spherical shells ($\Omega\simeq S^{2}\times[0,\delta]$), shown as a function of the reduced cavity size $R/\lambda_D$ (logarithmic scale). Although $\eta(R)\to 0$ as $R\to\infty$, finite-size deviations follow a topology-controlled hierarchy, $\text{spherical} > \text{cylindrical} > \text{planar}$. The dotted line indicates $\eta(R)=0$.}
	\label{fig:Two-Plates-charge profiles-and-eta}
\end{figure}

For the representative case shown in \Cref{fig:Two-Plates-charge profiles-and-eta}, we consider a $1{:}1$ electrolyte with bulk concentration $\rho_0=0.1\,\mathrm{M}$, temperature $T=298.15\,\mathrm{K}$, dielectric constant $\varepsilon=78.5$, and corresponding Debye length $\lambda_D=9.63\times10^{-10}\,\mathrm{m}$. The inner and outer surface charge densities on the shell's walls are fixed at $\sigma_{in}=\sigma_{out}=\sigma_0=0.0005\,\mathrm{C/m^2}$.

\Cref{fig:Two-Plates-charge profiles-and-eta} shows the behavior of $\eta(R)$ as a function of the reduced cavity size $R/\lambda_D$ for planar, cylindrical, and spherical geometries, where $\lambda_D$ is the Debye screening length, obtained from the analytical solutions of the linearized \cref{eq:PB}~\cite{Adrian-JML-2023,Lozada-Cassou_Symmetry_breaking_2025}. In all cases, $\eta(R)$ decreases monotonically and vanishes asymptotically as $R \to \infty$, indicating the restoration of local electroneutrality in the large-system limit, while global electroneutrality is preserved for all cavity sizes. 

However, at finite cavity size, the magnitude of the deviation exhibits a clear and systematic ordering across geometries. The strongest deviations occur in spherical confinement, followed by cylindrical and planar geometries, yielding the hierarchy
\begin{equation}
	\eta_{\mathrm{spherical}}(R) > \eta_{\mathrm{cylindrical}}(R) > \eta_{\mathrm{planar}}(R).
\end{equation}
This ordering is robust across variations in electrolyte concentration and surface charge density and reflects a structural property of the confining domain rather than by local geometric curvature alone.

The observed hierarchy originates from the topology of the confining domain. Spherical shells correspond to compact manifolds that enclose a finite volume, enforcing global electrostatic constraints through Gauss’ law. This global enclosure couples the entire charge distribution and enhances VLEC. In contrast, planar slits are non-compact and, although they also exhibit VLEC, its magnitude decreases monotonically with system size, with no enhanced nonlinear regime or minimum. Cylindrical geometries are only partially compact and therefore display an intermediate behavior between these two limits: their decay is stronger than in spherical shells, but weaker than in planar slits, and they develop a shallower minimum than the spherical case, as shown in Fig. 1.

Thus, while VLEC is a consequence of confinement, topology controls its scaling behavior and nonlinear response.

It is important to emphasize that these effects arise within Poisson--Boltzmann theory, without invoking ion-size effects or correlations beyond the mean-field description, which are typically associated with corrections to Poisson--Boltzmann theory~\cite{Netz-Orland-Beyond-PB-2000}. In the standard view, phenomena such as overcharging and charge reversal are attributed to configurational correlations associated with finite ion size and are therefore generally assumed to be absent in point-ion models. In contrast, the present results show that confinement-induced overcharging and charge reversal can emerge even for point ions as a consequence of global electrostatic constraints imposed by confinement. The self-consistent solution of the Poisson--Boltzmann equation couples the entire charge distribution through long-range Coulomb interactions, and this coupling becomes structurally constrained by the topology of the confining domain. As a result, nontrivial charge organization can arise within a mean-field description, reflecting global electrostatic constraints rather than the local correlation effects traditionally invoked to explain these phenomena~\cite{Attard_1996,Vlachy1989,Kjellander-charge-inversion-1998,Netz-Orland-Beyond-PB-2000,Deserno-2001,Grosse-2002,Jimenez_2004_Feb,Green-electroneutrality-JCP-2021,Henderson2012,Gonzalez-overcharging-cyl-macroions-2022,Lozada-Cassou_JML-2025}.

\textit{Conclusion.—}
In summary, we have shown that the finite-size violation of local electroneutrality in confined electrolytes is governed by the topology of the confining domain. The resulting hierarchy of deviations across spherical, cylindrical, and planar geometries demonstrates that charge redistribution under confinement is controlled by global electrostatic constraints, rather than by local geometric details or particle models. These findings establish topology as a fundamental organizing principle in the electrostatics of confined systems.

More broadly, the existence of VLEC reflects a global electrostatic constraint that determines the structure of both the confined region and the surrounding electrolyte. This constraint influences thermodynamic observables such as osmotic pressure, as well as adsorption phenomena in biological systems (e.g., cells and vesicles) and collective behavior in colloidal systems, highlighting the broad relevance of VLEC across soft condensed matter and biological contexts.

\textit{Acknowledgements.—}
The author gratefully acknowledges support from DGAPA-UNAM through PAPIIT Project No.~ IN108023. Useful discussions with collaborators and students over the years are also gratefully acknowledged.

%\bibliography{Topology-PRL}
%

\end{document}